\documentclass[trackchanges,twocolumn]{aastex7}

\usepackage{url}  
\usepackage{hyperref}
\usepackage{graphicx}
\usepackage{amsmath}
\usepackage[normalem]{ulem}
\usepackage{xcolor}
\usepackage{tabularx}
\usepackage{graphicx}    
\usepackage{subcaption}  
\usepackage{lineno}
\linenumbers

\usepackage{float}  

\begin{document}

\title{
An AI super-resolution field emulator for cosmological hydrodynamics: the Lyman-$\alpha$ forest.
}

\author[orcid=0000-0002-9140-3950]{Fatemeh Hafezianzadeh}
\affiliation{McWilliams Center for Cosmology, Department of Physics, Carnegie Mellon University, Pittsburgh, PA 15213, USA }
\email[show]{fhafezia@andrew.cmu.edu}  

\author[orcid=0009-0008-0370-6021]{Xiaowen Zhang} 
\affiliation{McWilliams Center for Cosmology, Department of Physics, Carnegie Mellon University, Pittsburgh, PA 15213, USA }
\email{fakeemail2@google.com}

\author[orcid=0000-0001-7899-7195]{Yueying Ni} 
\affiliation{Harvard-Smithsonian Center for Astrophysics, 60 Garden Street, Cambridge, MA 02138, USA}
\email{fakeemail2@google.com}

\author[orcid=0000-0003-0697-2583]{Rupert A. C. Croft} 
\affiliation{McWilliams Center for Cosmology, Department of Physics, Carnegie Mellon University, Pittsburgh, PA 15213, USA }
\email{fakeemail2@google.com}

\author[orcid=0000-0002-6462-5734]{Tiziana DiMatteo} 
\affiliation{McWilliams Center for Cosmology, Department of Physics, Carnegie Mellon University, Pittsburgh, PA 15213, USA }
\email{fakeemail2@google.com}

\author[orcid=0000-0001-7066-1240]{Mahdi Qezlou} 
\affiliation{Department of Astronomy, The University of Texas at Austin, 2515 Speedway Boulevard, Stop C1400, Austin, TX 78712, USA }
\email{fakeemail2@google.com}

\author[orcid=0000-0001-5803-5490]{Simeon Bird} 
\affiliation{Department of Physics and Astronomy, University of California Riverside, 900 University Ave, Riverside, CA 92521, USA }
\email{fakeemail2@google.com}

\begin{abstract}

We extend our super-resolution and emulation framework for cosmological dark matter simulations to include hydrodynamics. We present a two-stage deep learning model to emulate high-resolution (HR-HydroSim) baryonic fields from low-resolution (LR-HydroSim) simulations at redshift $z = 3$. The method takes as inputs an LR-HydroSim and the high-resolution initial conditions (HR-HydroICs). First, the model stochastically generates high-resolution baryonic fields from the LR-HydroSim.  Second, a deterministic emulator refines these fields using HR-HydroICs to  reconstruct small-scale structures including displacement, velocity, internal energy, and gas/star classification. Trained on paired low- and high-resolution simulations produced with \texttt{MP-Gadget}, the model captures small-scale structures of the intergalactic medium and 
observables down to the 100 kpc pressure smoothing scale relevant to the Lyman-$\alpha$ forest. The model achieves subpercent error for overdensity, temperature, velocity, and optical depth fields, a mean relative error of 1.07\% in the large-scale flux power spectrum (\(k < 3 \times 10^{-2}\ \mathrm{s/km}\)), and less than 10\% error in the flux probability distribution function. Notably, the two-stage model reduces the compute time by a factor of $\sim$450 compared to full smoothed particle hydrodynamics at the same resolution. 
This work demonstrates the potential of this framework as a powerful and efficient tool for generating high-resolution fields offering fast and accurate alternatives to traditional cosmological hydrodynamic simulations and enabling large-volume mock datasets for next-generation cosmological surveys.
\end{abstract}



\section{Introduction} 

A key element in studying galaxy formation and evolution is tracing the distribution and evolution of baryonic matter in the Universe—particularly the interplay between gas and stars. Cosmological hydrodynamical simulations are a central methodology employed for this purpose, where gravitational dynamics are coupled with fluid equations to model the complex behavior of cosmic gas within an expanding Universe \citep{vogelsberger2020}. Several state-of-the-art hydrodynamic simulation codes, including  \textsc{Ramses} \citep{teyssier2002ramses}, \textsc{Gadget}\citep{springel2005gadget}, \textsc{MP-Gadget} \citep{mpgadget},  \textsc{Enzo} \citep{bryan2014enzo}, and \textsc{Gizmo} \citep{hopkins2015gizmo}, have been developed to simulate a wide range of baryonic processes such as gas cooling, star formation, and feedback from stars and active galactic nuclei. However, these simulations are computationally expensive—often the most resource-intensive among cosmological simulations—making it challenging to generate large volumes or statistically significant ensembles, especially when high spatial resolution is required.

One of the most powerful applications of cosmological hydrodynamics is the modeling of the Lyman-$\alpha$ (Ly$\alpha$) forest—a series of absorption features observed in the spectra of distant quasars, caused by intervening neutral hydrogen in the intergalactic medium (IGM). As Ly$\alpha$ photons travel through the IGM, they are absorbed at specific redshifts corresponding to neutral hydrogen along the line of sight. The resulting absorption pattern is encoded in the transmitted flux, defined as $F = e^{-\tau}$, where $\tau$ is the Ly$\alpha$ optical depth. Since $\tau$ is proportional to the neutral hydrogen density, the transmitted flux provides a direct probe of the thermal and density structure of the IGM. This makes the Ly$\alpha$ forest a powerful tool for constraining cosmological parameters and the thermal history of the Universe \citep{lyaforest_review,mcquinn2016,bird2023priya, fernandez2024cosmological}

Observationally, Ly$\alpha$ forest data span a range of instrumental resolutions—from large spectroscopic surveys such as SDSS/BOSS \citep{dawson2013} and DESI \citep{desi2016}, with resolving powers of $R \sim 2000$--$5000$ (tens to hundreds of km/s in velocity), to targeted high-resolution studies like \citet{Day2019} and \citet{walther2019new}, which reach velocity resolutions of a few km/s. These observations probe scales down to the pressure smoothing scale of the gas—a physical limit set by the interplay of thermal pressure and gravitational collapse. However, the more fundamental modeling challenge arises from the highly non-linear relationship between gas density, temperature, velocity gradients, and Ly$\alpha$ absorption \citep{hui1997equation}. As a result, accurate predictions at any scale require simulations that resolve the pressure smoothing scale to faithfully capture the complex transformation between baryonic structure and transmitted flux \citep{peeples2010pressure}.

In order to accurately model the Ly$\alpha$ forest, simulations must resolve small-scale, non-linear baryonic structures, including filamentary gas and shocked regions in the IGM. High-resolution simulations are thus required, with typical line-of-sight resolutions on the order of a few kilometers per second \citep{viel2004inferring,bolton2009resolving,viel2013cosmology,lukic2015lyman}. Zoom–in approaches have been developed to explore the parameter space efficiently while maintaining this resolution \citep{onorbe2014zoom}, and large grids of high-resolution runs have been used to constrain the IGM thermal state and pressure smoothing scale \citep{walther2019new}. However, achieving such resolution across cosmologically large volumes remains computationally expensive in both runtime and memory.

Beyond full hydrodynamical simulations, several approximate methods have been developed to efficiently model the Ly$\alpha$ forest. Hydro-particle-mesh (hydro-PM) methods incorporate pressure effects into N-body simulations \citep{gnedin1998}, while LyMAS maps dark matter fields to flux using probability distributions calibrated on hydrodynamic runs \citep{peirani2014lymas, peirani2022}. Other approaches use Fluctuating Gunn–Peterson Approximation (FGPA) based models applied to dark matter-only simulations, often with added temperature-density relations \citep{sorini2016}. There are also recent promising models based on perturbation theory for larger scale flux statistics \citep{Roger2024eft}, and non-local FGPA extensions have been proposed to improve agreement with simulations \citep{sinigaglia2024field}.

Artificial intelligence is increasingly transforming cosmological research, and the study of the Lyman-$\alpha$ forest is no exception. A variety of AI-driven approaches have been proposed to reduce the computational cost associated with high-resolution simulations. In particular, emulator-based methods have been developed to interpolate summary statistics—such as the one-dimensional (1D) Ly$\alpha$ flux power spectrum (P1D)—across cosmological and astrophysical parameter spaces using techniques like Gaussian processes~\citep{bird2019} and neural networks~\citep{cabayol2023neural}. These models enable rapid and efficient parameter inference, making them valuable tools for exploring large cosmological model spaces.

To go beyond flux prediction, other models leverage deep learning to infer gas properties directly from dark matter fields. For example, \textsc{Ly$\alpha$NNA} uses ResNet architectures to extract thermal parameters from mock spectra and demonstrates improved performance over traditional power spectrum or PDF-based methods but just under ideal, noiseless conditions~\citep{nayak2024lyalphanna}. Similarly, \textsc{LyAl-Net} applies a 3D U-Net to map dark matter densities to temperature, neutral hydrogen, and flux~\citep{boonkongkird2023lyalnet}, building on earlier frameworks like \textsc{LyMAS}~\citep{peirani2014lymas}. \textsc{EMBER}, which combines a U-Net with a Wasserstein GAN, generates high-resolution gas and H\,\textsc{i} maps from dark matter-only simulations with 10\% accuracy ~\citep{bernardini2022ember}. Other approaches include U-Net architectures tailored for specific feedback processes to study specific physical processes. For example, ~\citep{list2019black} applied a U-Net to model the impact of dark matter annihilation on the intergalactic medium, diffusion-based models for baryonic field super-resolution~\citep{kodi2020super}, convolutional neural networks for rapid Ly$\alpha$ flux prediction~\citep{harrington2022fast}, and models trained to reconstruct Ly$\alpha$ fields from low-resolution hydrodynamical simulations~\citep{jacobus2023reconstructing}.

Despite their diversity, these models often focus on narrow tasks, lack generalizability across conditions, or fall short in predictive accuracy—highlighting the need for more robust and comprehensive solutions. While such AI-driven strategies have enabled rapid exploration of complex cosmological models, they typically do not address the challenge of enhancing resolution from coarse data. A key limitation is that most models either target isolated field quantities or operate at low spatial resolution, rather than producing detailed, high-resolution baryonic fields from limited inputs.The ability to perform super-resolution—reconstructing fine-grained physical structures such as gas density, temperature, and velocity at fixed cosmology—offers a promising path toward creating high-fidelity mock simulations beyond what is computationally feasible with traditional methods.

In this work, we address this challenge by developing a novel emulator framework HydroEmu, that extends our previous super-resolution models~\citep{li2021ai,ni2021ai,zhang2024ai,zhang2025ai} to hydrodynamic cosmological simulations. Our model is trained to predict the density, velocity, and internal energy fields of both gas and stars\footnote{For stars, however, the internal energy was not predicted but instead assigned a fixed value.}, while also classifying particles as gas or stellar components. The stellar field is included primarily to ensure that stellar particles— which do not contribute to Lyman-$\alpha$ absorption—are correctly identified and excluded from the flux calculations. Conditioned on high-resolution initial conditions and trained adversarially, the emulator achieves 0.1-10\% accuracy, depending on the metric.  This represents a significant advance toward enhancing computationally intensive baryonic simulations with AI-based models that can extend their effective resolution and dynamic range, capable of producing full-field outputs suitable for downstream analysis. 

Training data are generated using the \texttt{MP-Gadget} hydrodynamical code \citep{mpgadget}, focusing on gas and star particles to isolate the baryonic contributions to Ly$\alpha$ absorption. Dark matter components are omitted for simplicity. The fidelity of the model is assessed using the \texttt{fake\_spectra} tool \citep{fake_spectra_tool} to generate synthetic Ly$\alpha$ absorption spectra. Comparisons are made between HydroEmu predictions, HR-HydroSim outputs, and observational measurements of flux and optical depth.

In the following sections, we evaluate HydroEmu through a series of tests: visual comparisons of field morphology, quantitative field-level analysis across multiple physical quantities, recovery of the temperature--density relation, and assessments of flux statistics using the power spectrum and PDF. This multi-faceted evaluation benchmarks the emulator against HR-HydroSim and observational data.

\section{Method} 
\subsection{Dataset}
A dataset was constructed using cosmological hydrodynamical simulations carried out with \texttt{MP-Gadget}\footnote{\url{https://github.com/MP-Gadget/MP-Gadget}}, a code that combines gravitational dynamics and smoothed particle hydrodynamics (SPH). Gravitational forces are computed using the TreePM algorithm \citep{bagla2002}, which splits the calculation into long-range interactions solved via a particle-mesh (PM) method in Fourier space, and short-range forces evaluated using a hierarchical tree (octree) structure. The SPH implementation models baryonic processes including radiative cooling, photoionization heating from a uniform ultraviolet background \citep{haardt2012}, and a simplified prescription for star formation based on density thresholds \citep{springel2003}. These simulations adopt the “quick-Ly$\alpha$” approximation \citep{viel2004inferring}, in which gas particles that exceed a specified density are converted into star particles, thereby suppressing high-density gas clumping and improving computational efficiency while preserving the neutral hydrogen distribution relevant for Lyman-$\alpha$ forest analyses. This approximation does not model galaxies, stars, or black holes in a physically realistic manner, and the resulting star particles should not be interpreted as resolved stellar systems. The simulations share their codebase with recent large-scale projects such as the ASTRID simulations \citep{ni2022astrid,bird2022astrid}, but are specifically optimized for modeling the intergalactic medium and transmitted flux in the Lyman-$\alpha$ forest at high redshift.

Each simulation consisted of a LR-HydroSim and a HR-HydroSim run within a cubic volume of $50\, h^{-1}\,\mathrm{Mpc}$ per side. The LR-HydroSim simulations employed $64^3$ particles, while the HR-HydroSim counterparts used $512^3$ particles. It should be noted that the resolution of the HR-HydroSim runs is not sufficient to fully resolve the Lyman-$\alpha$ forest with high accuracy; however, this is acceptable for the purpose of demonstrating the methodology. Initial conditions were generated at redshift $z = 99$ using first-order Lagrangian perturbation theory. The linear power spectrum used for initialization was computed with \texttt{CLASS} \citep{lesgourgues2011}. All simulations followed a WMAP9 cosmology \citep{hinshaw2013}, with parameters $\Omega_m = 0.2814$, $\Omega_\Lambda = 0.7186$, $\Omega_b = 0.0464$, and $h = 0.697$. Gas particles were initialized with a temperature of 270~K, and cooling was allowed down to a floor temperature of 100~K. Star formation uses the quick-Ly$\alpha$ approximation, which converts gas to stars when it reaches an overdensity greater than $10^3$ times the critical density. This model removes high-density gas from the simulation, preventing prevent it from dominating the computational cost of the simulation, focusing simulation resources on low-density gas associated with the Lyman-$\alpha$ forest \citep{viel2004inferring}.

In total, 20 simulation pairs were generated, with 16 used for training and validation and 4 reserved for testing. Each simulation was evolved to redshift $z = 3$, where physical fields—such as gas density, temperature, velocity, and optical depth—were extracted along 3600 regularly spaced sightlines aligned with the $y$-axis. Mock absorption spectra were generated using the \texttt{fakespectra} code\footnote{\url{https://github.com/sbird/fake_spectra}}, which computes optical depth by integrating neutral hydrogen absorption along each sightline, accounting for thermal broadening, peculiar velocities, and redshift-space distortions. The code also outputs interpolated gas properties and the transmitted flux, $F = \exp(-\tau)$, at each pixel. Sightlines consist of 540 pixels sampled at $10 \mathrm{km/s}$ resolution, covering a velocity range of $5400 \mathrm{km/s}$—sufficient to resolve the small-scale structure of the Lyman-$\alpha$ forest.

\subsection{Model Architecture}

\begin{figure*}[ht!]
\centering
\includegraphics[width=\textwidth]{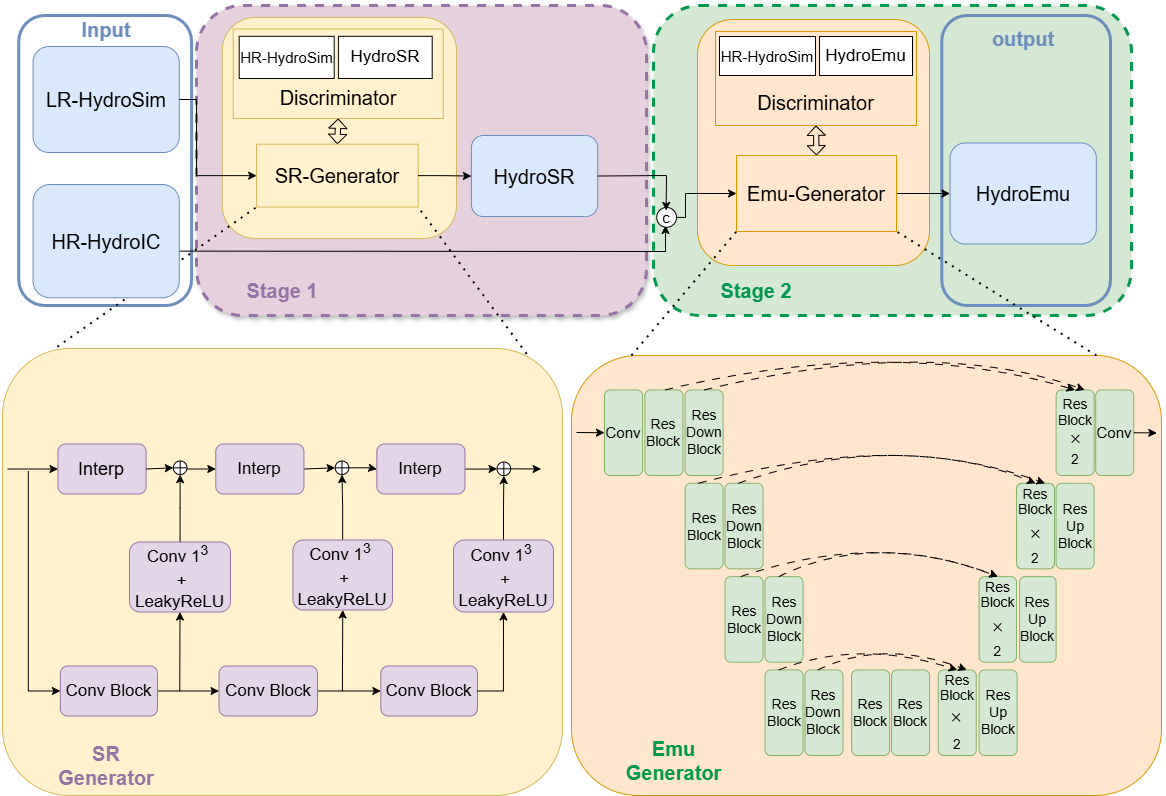}
\caption{Schematic overview of the proposed two-stage framework. The upper panel illustrates the overall model architecture and training process, including Stage 1 (left, purple box) and Stage 2 (right, green box). The lower panels show the detailed generator architectures for each stage: the SR-Generator (yellow box) used in Stage 1 and the Emu-Generator (orange box) used in Stage 2.}
\label{fig:model_schematic}
\end{figure*}
Generative Adversarial Networks (GANs) are a class of deep generative models introduced by \citet{goodfellow2014generative}. A GAN consists of two neural networks with distinct roles: a generator, which produces synthetic data, and a discriminator, which attempts to distinguish between real and generated samples. Each of these networks may adopt different architectural designs depending on the task. GAN-based models have demonstrated remarkable success in generating high-fidelity outputs, particularly in image synthesis and super-resolution applications~\citep{ledig2017photo, wang2018esrgan}.

One of the common architectural choices for the generator network in GANs is the U-Net~\citep{ronneberger2015u}. Originally developed for biomedical image segmentation, U-Net adopts an encoder-decoder structure with skip connections that bridge corresponding layers between the downsampling and upsampling paths. These skip connections help retain fine-grained spatial information lost during encoding, enabling accurate and detailed reconstructions. U-Nets have been widely used in image-to-image translation tasks~\citep{isola2017image}, including super-resolution, denoising, and emulation of physical fields in cosmology. In conditional settings(conditional U-Net), additional physical or contextual information can be appended to the input channels, allowing for more controlled and physically-informed generation.

In this work, a two-stage GAN-based framework is developed to emulate high-resolution baryonic fields relevant to the Lyman-$\alpha$ forest at redshift $z = 3$. Although both stages adopt the GAN paradigm, they differ in architecture and purpose. The first stage, HydroSR, is a stochastic super-resolution model following the architecture proposed by \citet{ni2021ai}, which generates reasonable high-resolution fields from low-resolution inputs. The second stage, HydroEmu, is a deterministic emulator based on the conditional U-Net architecture introduced in \citet{zhang2025ai}. It is trained to refine the output of HydroSR and accurately reconstruct the small-scale structure present in high-resolution hydrodynamic simulations.

The model architectures used in both stages closely follow the designs presented in \citet{ni2021ai} for HydroSR and \citet{zhang2025ai} for HydroEmu, with only minor modifications to support additional input channels—such as internal energy and a binary gas/star label—which will be discussed in detail later. A schematic illustration of the full framework, including the generator networks for HydroSR and HydroEmu, is shown in Figure~\ref{fig:model_schematic}.

\textbf{Stage 1: HydroSR model.}  
The generator architecture employed in HydroSR was adapted from a model originally designed for super-resolving dark matter displacement and velocity fields \citep{ni2021ai}, and was extended to handle baryonic quantities. The network receives as input a set of 3D particle-based fields with eight channels that collectively encode gas and stellar components: three channels for displacements, three for velocity, one for internal energy, and one binary label that differentiates gas from stars.

As illustrated in Figure~\ref{fig:model_schematic} (lower panel, yellow box), the generator features a hierarchical structure in which information is processed across multiple spatial resolutions. The lower branch of the network operates in a latent space through a cascade of convolutional layers, while the upper branch projects intermediate representations into the target output space at successive resolutions. Each scale includes a projection step, followed by trilinear interpolation and accumulation of outputs across levels. The generator uses three-dimensional convolutional kernels, with Leaky ReLU activations (negative slope 0.2) applied after each layer.

The discriminator adopts a fully convolutional architecture similar to PatchGAN \citep{isola2017image}, incorporating residual connections as described in \citet{li2021ai}. It evaluates local patches of the field and estimates the Wasserstein distance used in the adversarial objective. The HydroSR model is trained using a total loss that combines (i) a supervised Lagrangian loss on particle-based fields including displacement, velocity, internal energy, and gas/star labels, (ii) a supervised Eulerian loss computed after mapping particle fields to grid space via cloud-in-cell deposition (including density and energy density), and (iii) an adversarial loss using the WGAN-GP\citep{gulrajani2017improved} formulation. The adversarial discriminator is trained conditionally on the input fields and encourages the generator to produce physically plausible high-resolution outputs.
\begin{equation}
\mathcal{L}_{\text{total}} =
\mathcal{L}_{\text{Lag}}^{\text{MSE}} +
\mathcal{L}_{\text{Eul}}^{\text{MSE}} +
\lambda_{\text{adv}} \, \mathcal{L}_{\text{adv}}^{\text{WGAN-GP}},
\label{eq:loss}
\end{equation}
where both $\mathcal{L}_{\text{Lag}}^{\text{MSE}}$ and $\mathcal{L}_{\text{Eul}}^{\text{MSE}}$ are computed using the standard L2 mean squared error (MSE):
\begin{equation}
\mathcal{L}^{\text{MSE}} = \frac{1}{N} \sum_{i=1}^{N} \left( \hat{x}_i - x_i \right)^2,
\end{equation}
and $\lambda_{\text{adv}}$ is a weighting factor for the adversarial loss term. The WGAN-GP loss \citep{gulrajani2017improved} is given by:
\begin{align}
\mathcal{L}^{\text{WGAN-GP}} =\;& 
\mathbb{E}_{\boldsymbol{\ell}, \mathbf{z}} \left[ D(\boldsymbol{\ell}, G(\boldsymbol{\ell}, \mathbf{z})) \right]
- \mathbb{E}_{\boldsymbol{\ell}, \mathbf{h}} \left[ D(\boldsymbol{\ell}, \mathbf{h}) \right] \nonumber \\
&+ \lambda \, \mathbb{E}_{\boldsymbol{\ell}, \mathbf{h}} \left[ \left( \left\| \nabla_{\mathbf{i}} D(\boldsymbol{\ell}, \mathbf{i}) \right\|_2 - 1 \right)^2 \right]
\label{eq:WGAN}
\end{align}
where $\boldsymbol{\ell}$ denotes the input, $\mathbf{z}$ is a latent noise vector, $\mathbf{h}$ is a real HR-HydroSim sample, and $G(\boldsymbol{\ell}, \mathbf{z})$ is a fake generated sample. The first two terms correspond to the Wasserstein distance between the real and generated distributions, while the third enforces the gradient penalty to satisfy the 1-Lipschitz constraint. The coefficient $\lambda$ was set to 10, as is commonly adopted in similar contexts. During training, the discriminator is optimized to minimize all three terms in $\mathcal{L}_{\text{WGAN-GP}}$, while the generator is trained to maximize only the first term, thereby learning to produce realistic outputs that can fool the discriminator. Once the adversarial training converges and the total loss $\mathcal{L}_{\text{total}}$ is minimized, the generator (i.e., HydroSR or HydroEmu) is capable of producing high-fidelity outputs indistinguishable from real data. At inference time, the discriminator is discarded, and only the trained generator is used to generate high-resolution predictions.

\textbf{Stage 2: HydroEmu model.}

The purpose of this stage, and the use of HydroEmu specifically, is to refine the output generated by the stochastic HydroSR model so that it closely matches the corresponding HR-HydroSim realization, conditioned on the high-resolution initial conditions (HR-HydroICs). A deterministic emulator was employed using the exact architecture proposed in \citet{zhang2025ai}, with minor modifications made to support eight input channels.

As illustrated in Figure~\ref{fig:model_schematic} (lower panel, orange box), the generator adopts a U-Net-like architecture composed of residual blocks with group normalization and \textsc{SiLU} \citep{elfwing2018sigmoid} activation. The input to the network is constructed by concatenating the 8-channel output of the HydroSR model with 8 additional channels derived from the HR-HydroICs. These include gas positions, velocity components, internal energy, and a binary gas/star label that is uniformly set to ``gas,'' based on the assumption that at redshift $z = 99$ all particles are in the gas phase. The resulting 16-channel volumetric input is processed through a hierarchy of downsampling and upsampling blocks, allowing the network to reconstruct small-scale baryonic structures conditioned on both coarse output and fine initial conditions.

The HydroEmu model is trained using the same composite loss function as HydroSR (Equation~\ref{eq:loss}), consisting of supervised Lagrangian losses on particle-based displacement, velocity, internal energy, and gas/star labels, as well as an adversarial loss term based on the WGAN-GP formulation (Equation~\ref{eq:WGAN}). The only distinction is that, in HydroEmu, the Eulerian loss term is applied exclusively to the density field.

\subsection{Error Metrics}
\label{sec:metrics}

To quantify the agreement between HydroEmulator predictions and HR-HydroSim outputs, two standard error metrics were employed: the root mean square error (RMSE) and the normalized root mean square error (NRMSE).

The RMSE is defined as
\begin{equation}
    \mathrm{RMSE} = \sqrt{\frac{1}{N} \sum_{i=1}^{N} (x_i - \hat{x}_i)^2},
\end{equation}
where $x_i$ and $\hat{x}_i$ denote the true and predicted values, respectively, and $N$ is the total number of data points.

To enable scale-independent comparison across different physical fields, the RMSE was normalized by the dynamic range of the target quantity:
\begin{equation}
    \mathrm{NRMSE} = \frac{\mathrm{RMSE}}{x_{\text{max}} - x_{\text{min}}} \times 100\%,
\end{equation}
where $x_{\text{max}}$ and $x_{\text{min}}$ represent the maximum and minimum values of the HR-HydroSim reference field. This normalization allows direct comparison of predictive accuracy across quantities with different physical units and scales.

\section{Results}

The performance of HydroEmu was evaluated through comparisons with HR-HydroSim, considering both field-level physical quantities and statistical observables relevant to the Lyman-$\alpha$ forest. The following subsections present different aspects of the evaluation to comprehensively assess HydroEmu's accuracy.

\subsection{Visual Comparison}

A direct comparison between LR-HydroSim, HR-HydroSim, and HydroEmu was carried out to evaluate the HydroEmulator's capability in recovering the spatial and thermal structure of the IGM. As illustrated in Figure~\ref{fig:fullbox_zoom_flux}, LR-HydroSim exhibits a smoothed and less structured representation of the cosmic web, with significantly reduced small-scale features in both density and temperature. In contrast, HR-HydroSim displays well-defined filamentary structures and strong thermal contrast, capturing the complexity of the IGM on small scales.

HydroEmu closely reproduces the fine-grained morphology and thermal gradients seen in HR-HydroSim. The filamentary networks and dense knots of gas, largely absent in LR-HydroSim, are accurately recovered in HydroEmu. Furthermore, the transmitted Lyman-$\alpha$ flux extracted along the representative sightline shows strong agreement between HydroEmu and HR-HydroSim, with matching absorption features and amplitude variations. This level of visual similarity suggests that HydroEmu successfully reconstructs the non-linear structures essential for modeling the Lyman-$\alpha$ forest, effectively transforming the coarse LR-HydroSim input into a physically detailed output comparable to HR-HydroSim.

\subsection{Thermal State}

\begin{figure*}[ht!]
\centering
\includegraphics[width=\textwidth]{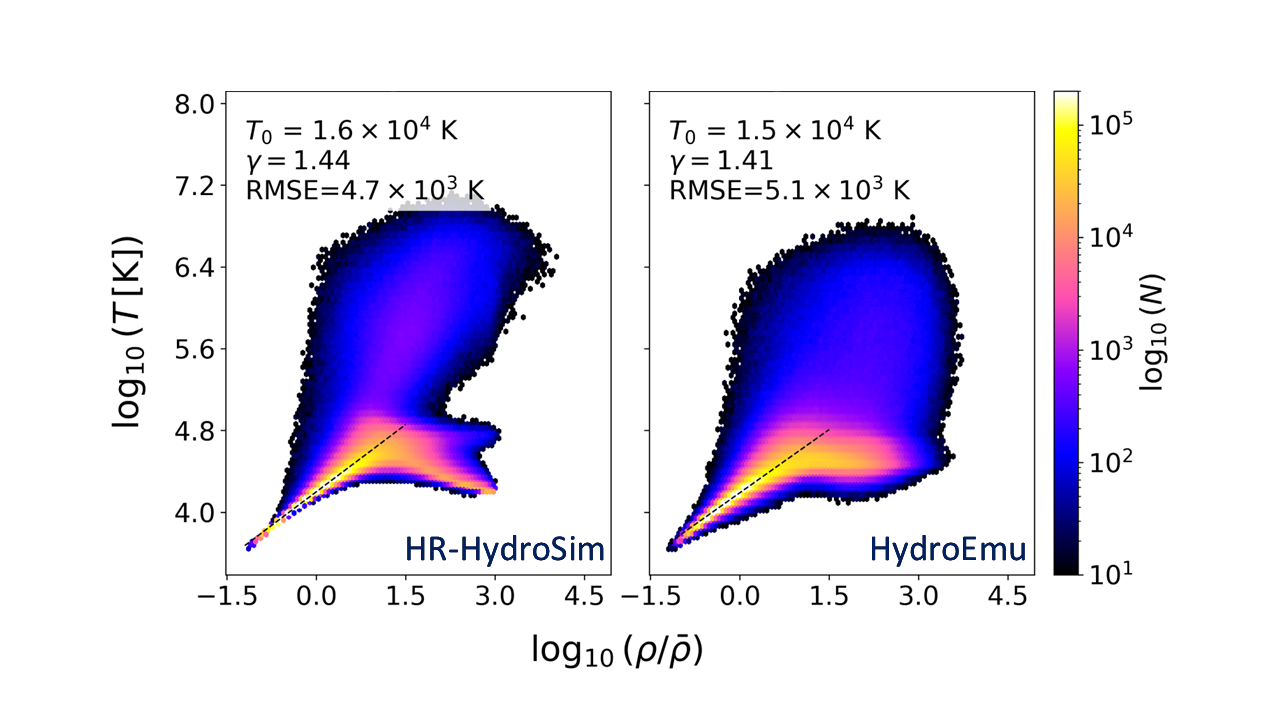}
\caption{Phase-space distribution of gas in the temperature--density plane for the HR-HydroSim (left) and the HydroEmu prediction (right). The dashed black line indicates the best-fit power-law relation to the IGM temperature-density relation, with $T_0$ and $\gamma$ denoting the temperature at mean density and fit parameter. The color scale shows the logarithmic number of gas elements per bin.}
    
\label{fig:phase_space}
\end{figure*}

The temperature-density relation (TDR) of the intergalactic medium was used as an additional diagnostic to evaluate the HydroEmulator’s physical accuracy. Figure~\ref{fig:phase_space} presents two-dimensional histograms of $\log_{10}(T\, [\mathrm{K}])$ versus $\log_{10}(\rho / \bar{\rho})$ for both the HR-HydroSim (left) and HydroEmu output (right). A power-law model of the form $T = T_0 (\rho / \bar{\rho})^{\gamma - 1}$ was fitted to the data in each case to extract the thermal state parameters. The fit was restricted to the well-defined TDR regime, specifically the region where $-1.0 < \log_{10}(\rho/\bar{\rho}) < 1.0$ and $0.1 < \log_{10}(T/\mathrm{K}) < 5$, ensuring that the analysis excluded shock-heated or high-density gas outside the typical Lyman-$\alpha$ forest phase.

For the HR-HydroSim, the best-fit parameters were found to be $T_0 = 1.6 \times 10^4$~K and $\gamma = 1.44$, while the HydroEmulator yielded $T_0 = 1.5 \times 10^4$~K and $\gamma = 1.41$. Here, $T_0$ represents the temperature at mean hydrogen density. The RMSE, reported in the figure, quantifies the scatter of temperature values from the fitted power-law model within the selected TDR regime. For the HydroEmulator, the RMSE was $5.1 \times 10^3$~K. When normalized by a dynamical range of $10^7$~K, the corresponding NRMSE was approximately 0.051\%. For comparison, the RMSE of the HR-HydroSim was $4.7 \times 10^3$~K, resulting in an NRMSE of about 0.047\%. These results demonstrate that the HydroEmulator not only reproduces the overall thermal structure of the intergalactic medium but also captures the power-law scaling behavior that governs the low-density IGM, with a normalized scatter of less than 0.1\%.

\subsection{Field-Level Comparison Along a Sightline}

\begin{figure*}[ht!]
\centering
\includegraphics[width=\textwidth]{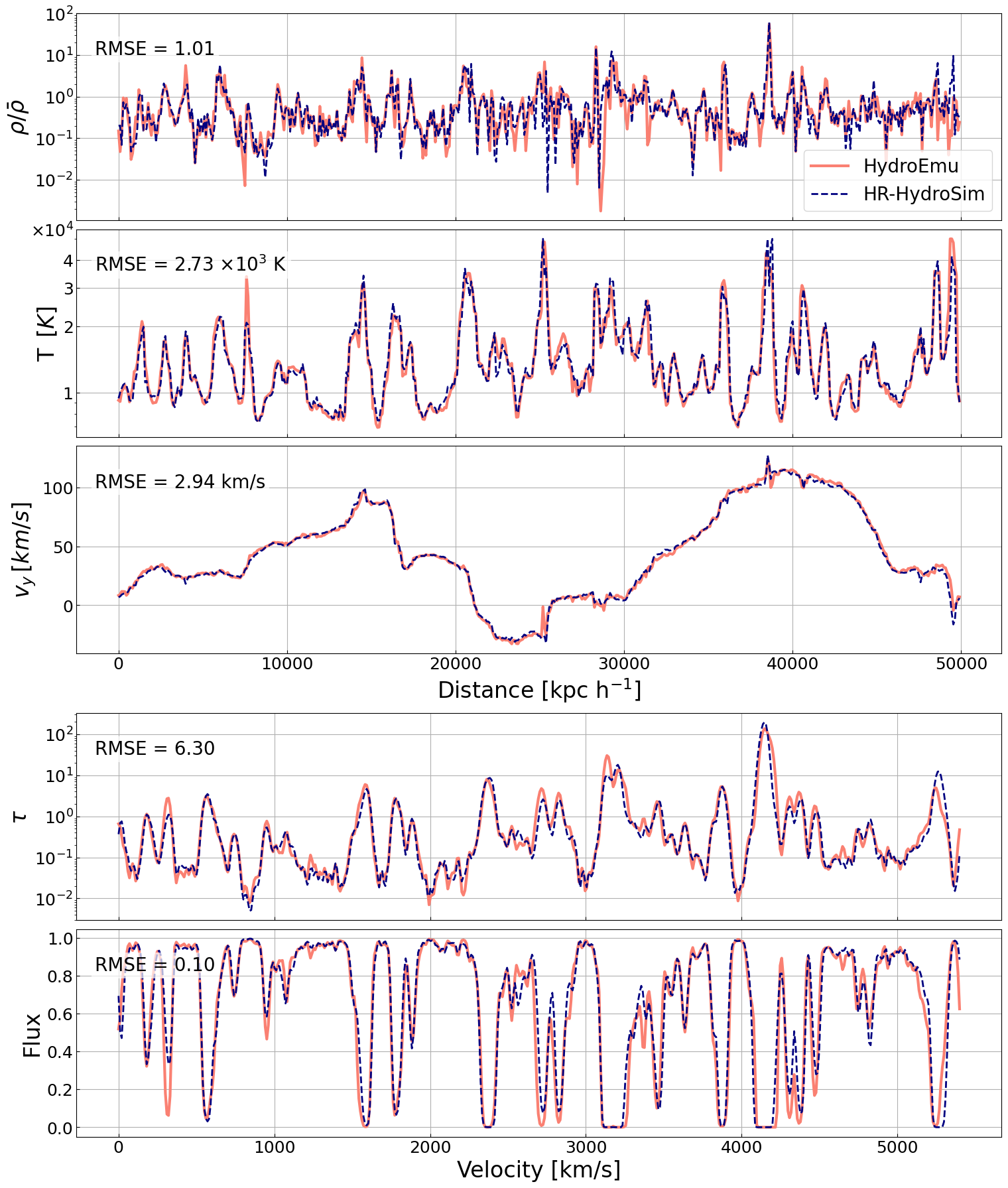}
\caption{
Comparison between the HR-HydroSim simulation and the HydroEmu prediction along a randomly selected sightline. The top three panels show the hydrogen overdensity $\rho / \bar{\rho}$, temperature $T$, and line-of-sight velocity $v_y$ in real (comoving) space, while the bottom two panels show the optical depth $\tau$ and transmitted flux in redshift (velocity) space. Solid red lines correspond to the HydroEmu predictions, while dashed dark blue lines represent the HR-HydroSim simulation results.}
\label{fig:sightline_comparison}
\end{figure*}

To evaluate HydroEmu at the field level, Figure~\ref{fig:sightline_comparison} presents a comparison between HydroEmu and HR-HydroSim across five different physical fields—density, temperature, velocity, optical depth, and transmitted flux—along a randomly selected sightline. The RMSE were computed for each field to quantify the agreement. The RMSE values are 1.01 for density ($\rho/\bar{\rho}$), $2.73 \times 10^3$~K for temperature, 2.94~km/s for velocity, 6.3 for optical depth, and 0.10 for flux. Corresponding NRMSE values are 1.67\%, 6.36\%, 1.84\%, 3.21\%, and 10.0\%, respectively, indicating high fidelity between HydroEmu predictions and HR-HydroSim ground truth, particularly for the density, temperature, and velocity fields.

\begin{figure*}[ht!]
\centering

\begin{minipage}[b]{0.89\textwidth}
    \centering
    \begin{subfigure}[b]{0.32\textwidth}
        \centering
        \includegraphics[width=\linewidth]{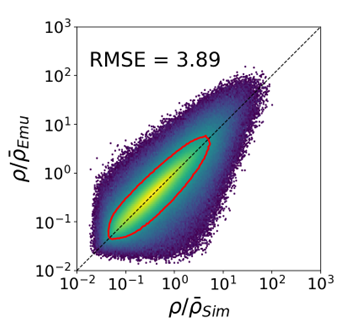}
        \caption*{(a) Overdensity}
    \end{subfigure}
    \begin{subfigure}[b]{0.32\textwidth}
        \centering
        \includegraphics[width=\linewidth]{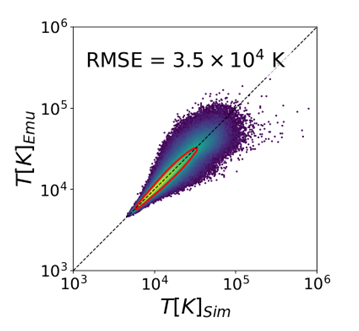}
        \caption*{(b) Temperature}
    \end{subfigure}
    \begin{subfigure}[b]{0.32\textwidth}
        \centering
        \includegraphics[width=\linewidth]{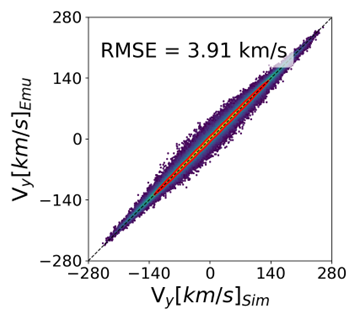}
        \caption*{(c) Velocity}
    \end{subfigure}

    \vspace{0.5cm}

    \begin{subfigure}[b]{0.32\textwidth}
        \centering
        \includegraphics[width=\linewidth]{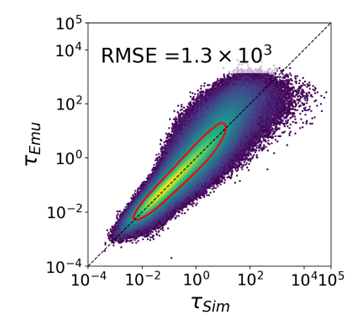}
        \caption*{(d) Optical Depth}
    \end{subfigure}
    \begin{subfigure}[b]{0.32\textwidth}
        \centering
        \includegraphics[width=\linewidth]{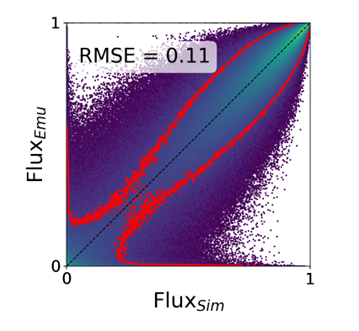}
        \caption*{(e) Flux}
    \end{subfigure}
    
\end{minipage}
\hfill
\begin{minipage}[b]{0.1\textwidth}
    \centering
    \includegraphics[height=11cm]{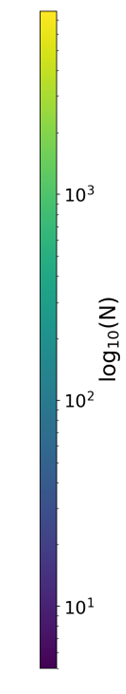} 
    
\end{minipage}

\caption{Comparison between HR-HydroSim outputs and HydroEmu predictions for key physical quantities in the Lyman-$\alpha$ forest. Each panel shows a 2D hexbin scatter plot of the HydroEmu prediction versus the HR-HydroSim ground truth for: (a) overdensity $\rho / \bar{\rho}$, (b) temperature $T\,[\mathrm{K}]$, (c) velocity component $V_y\,[\mathrm{km/s}]$, (d) optical depth $\tau$, and (e) transmitted flux. The dashed black line indicates one-to-one correspondence. RMSE values are shown in the top-left of each panel. Red contours enclose 90\% of the data.}
\label{fig:scatter-metrics}
\end{figure*}

To evaluate performance across the full simulation domain, and different simulation test set, the analysis was extended to a $60 \times 60$ grid of sightlines sampled from four simulation volumes. Pixel-wise comparisons between HydroEmu and HR-HydroSim are summarized in Figure~\ref{fig:scatter-metrics} using two-dimensional density histograms. Each panel corresponds to a different physical quantity, with the dashed diagonal line representing perfect agreement. RMSE and NRMSE values were computed for five physical quantities: overdensity, temperature, velocity, optical depth, and transmitted flux. The RMSE values are 3.89 for the overdensity, $3.5 \times 10^4$~K for temperature, 3.91~km/s for velocity, $1.3 \times 10^3$ for optical depth, and 0.11 for flux. To allow comparison across variables with different physical scales, each RMSE was normalized by the dynamic range of the corresponding HR-HydroSim field. The resulting NRMSE values are approximately 0.34\%, 0.28\%, 0.59\%, 0.33\%, and 10.83\% for the respective fields, confirming consistent performance across the test dataset.

The differences between the single-sightline and full-volume results can be attributed to the statistical variation in physical field values across the simulation. In particular, the temperature field exhibits significant variation across the dataset, with a broad dynamic range and a strongly skewed distribution concentrated around lower values. In contrast, the selected random sightline spans a comparatively narrow temperature range, resulting in a smaller denominator when computing NRMSE. Consequently, although the absolute RMSE for temperature was relatively low on this sightline, the NRMSE appeared comparatively larger. Similar effects apply to density, velocity, and optical depth, where a less representative local distribution leads to higher relative errors. Notably, the flux field—being bounded between 0 and 1 and derived through a nonlinear mapping from the other fields—exhibits consistent NRMSE values across both individual and full-sample evaluations, underscoring HydroEmu’s robust performance in predicting observable Lyman-$\alpha$ flux features.

\begin{figure*}[ht!]
    \centering
    \includegraphics[width=\textwidth]{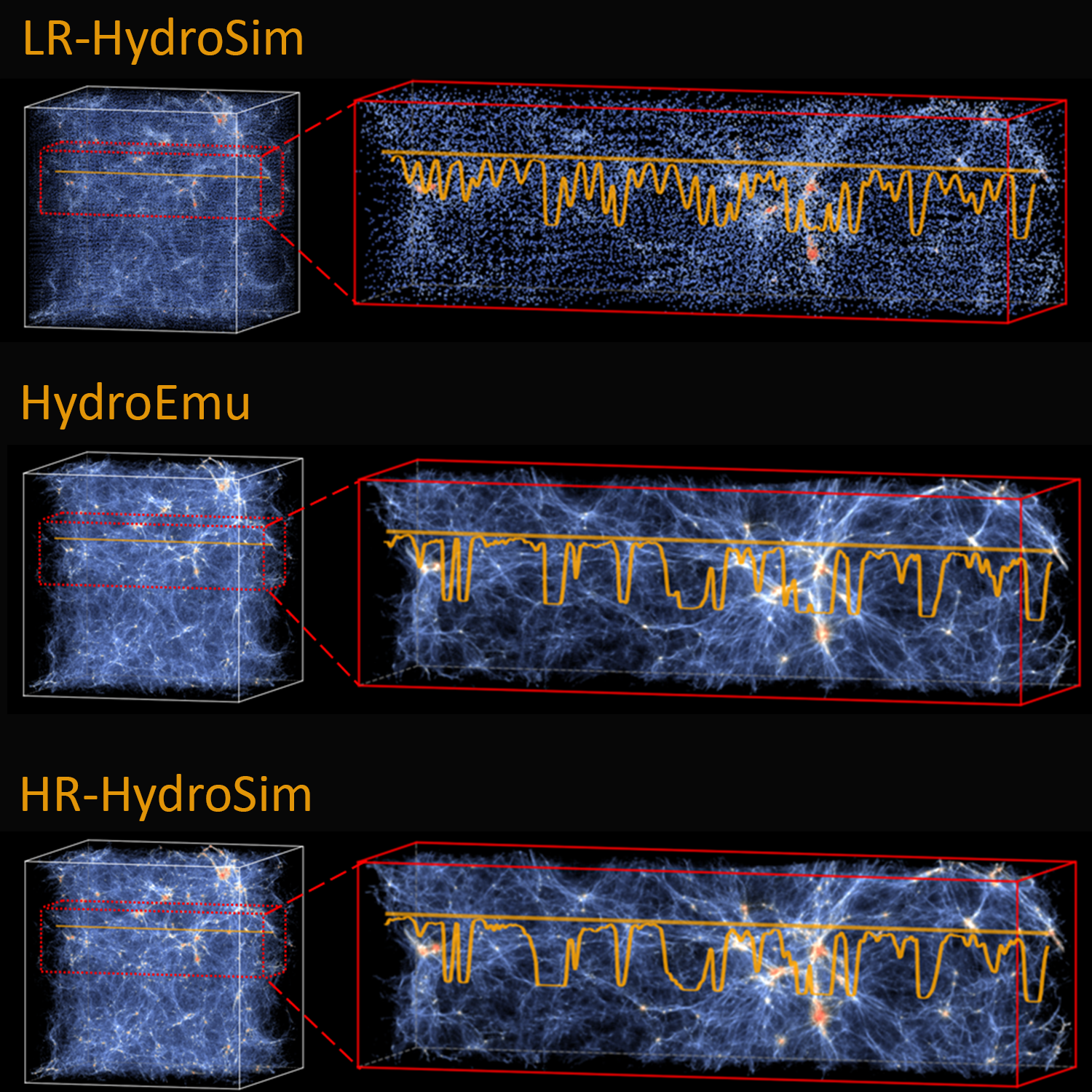}
    \caption{Comparison of gas distributions in the LR-HydroSim, HR-HydroSim, and HydroEmu simulations. The left column shows the full simulation volumes, where the color represents temperature and intensity indicates density. A representative sightline is highlighted by the red box and magnified in the right column. The corresponding Lyman-$\alpha$ flux profile along the sightline is overplotted in orange. The HydroEmulator prediction closely reproduces both the thermal structure and absorption features observed in the HR-HydroSim.}
    \label{fig:fullbox_zoom_flux}
\end{figure*}

These results indicate that HydroEmu accurately recovers the underlying IGM properties along individual sightlines. Compared to LR-HydroSim, as shown in Figure~\ref{fig:fullbox_zoom_flux}, HydroEmu provides a significantly more accurate reconstruction of the transmitted flux.

\subsection{Flux Statistics}

To provide a more comprehensive understanding of Lyman-$\alpha$ flux statistics, both one-point and two-point summary statistics are evaluated. In this analysis, the flux probability distribution function (PDF) is treated as a one-point statistic, while the one-dimensional flux power spectrum and flux decoherence are considered two-point statistics. These complementary diagnostics are used to assess the ability of HydroEmu to reproduce both the local distributional properties and the spatial correlations of the transmitted flux field. The emulator’s performance is validated through comparisons with HR-HydroSim outputs and observational datasets across both the flux PDF and power spectrum.

\subsection{Flux Decoherence Analysis}

To quantify the similarity between the predicted flux fields and those of the high-resolution simulation, we compute the Fourier-space cross-correlation coefficient $r(k)$ and use it to define the scale-dependent decoherence statistic:
\begin{equation}
    1 - r^2(k) = 1 - \left[ \frac{\operatorname{Re}\left( \langle \tilde{\delta F}_1(k) \tilde{\delta F}_2^*(k) \rangle \right)}{\sqrt{\langle |\tilde{\delta F}_1(k)|^2 \rangle \, \langle |\tilde{\delta F}_2(k)|^2 \rangle}} \right]^2.
\end{equation}
Here, $\tilde{\delta F}_i(k)$ denotes the Fourier transform of themean-normalized flux fluctuation, $\delta F_i(x) = F_i(x)/ \langle F_i(x) \rangle -1$, and the average is taken over multiple independent sightlines. This decoherence metric captures both amplitude and phase discrepancies between the compared fields.

\begin{figure}[ht!]
    \centering
    \includegraphics[width=\columnwidth]{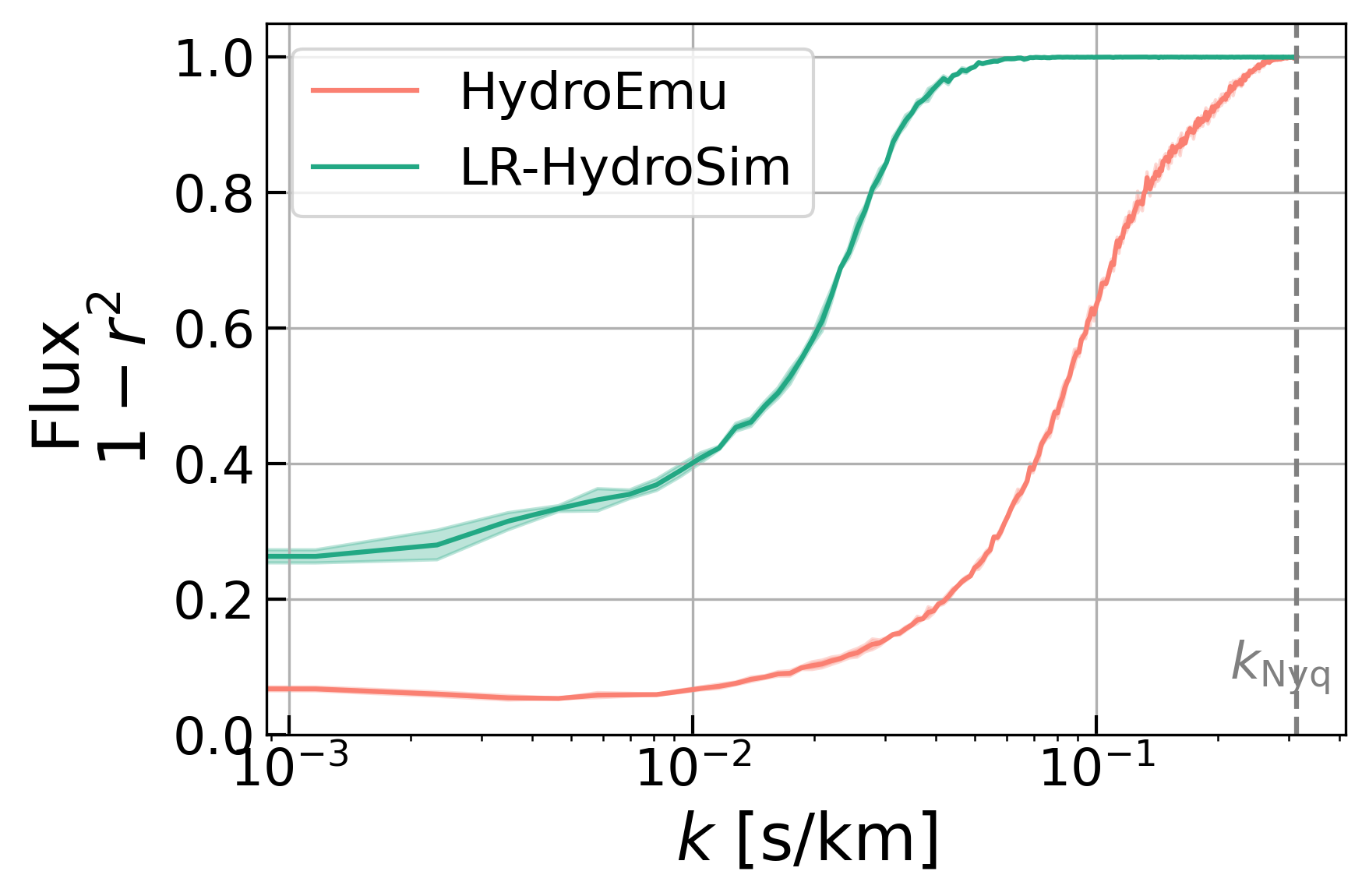}
    \caption{Scale-dependent flux decoherence, $1 - r^2(k)$, between the high-resolution simulation and each of the emulator (HydroEmu, red) and low-resolution simulation (LR-HydroSim, green). Shaded regions show the standard deviation across four test set. The vertical dashed line marks the Nyquist frequency $k_{\rm Nyq}$.}
    \label{fig:flux_decoherence}
\end{figure}

We apply this analysis to assess how closely the HydroEmu and the LR-HydroSim reproduce the flux statistics of the HR-HydroSim. For each model, the quantity $1 - r^2(k)$ is computed across four independent simulation sets, and the mean and standard deviation are shown in Figure~\ref{fig:flux_decoherence}.

The Nyquist wavenumber $k_{\rm Nyq}$ is defined as
\begin{equation}
    k_{\rm Nyq} = \frac{\pi}{\Delta v},
\end{equation}
where $\Delta v$ is the velocity bin size of the flux spectra. Here $\Delta v= 10$ km/s for LR-HydroSim, HR-HydroSim and HydroEmu). This corresponds to the maximum spatial frequency that can be resolved on a grid with pixel spacing $\Delta v$. All modes with $k > k_{\rm Nyq}$ are not physically meaningful and are excluded from the analysis.

As shown in Figure~\ref{fig:flux_decoherence}, HydroEmu consistently maintains a higher degree of coherence with the high-resolution flux field across all spatial scales. On large scales (\( k \lesssim 10^{-2} \, \mathrm{s/km} \)), HydroEmu exhibits lower flux decoherence compared to LR-HydroSim, indicating better alignment in both amplitude and phase of the predicted flux fluctuations. While both models begin with moderate decoherence at large scales, the divergence becomes increasingly pronounced with growing wavenumber. At \( k = 0.1 \, \mathrm{s/km} \)—the smallest scale at which flux statistics are reliably measured in observational studies \citep{Day2019,walther2019new}—the decoherence value for HydroEmu remains around 0.6, whereas the LR-HydroSim prediction approaches saturation with \( 1 - r^2(k) \approx 1 \). This highlights HydroEmu’s enhanced capability in recovering small-scale flux structures that are unresolved in the original low-resolution simulations.

Unlike the flux power spectrum, which is sensitive only to the amplitude of fluctuations, the decoherence statistic captures both amplitude and phase alignment. This makes it a more stringent and informative measure of the fidelity with which models like HydroEmu reproduce the detailed structure of the Lyman-\(\alpha\) forest. Although both HydroEmu and HR-HydroSim flux fields are rescaled to match the same mean flux, the decoherence statistic \(1 - r^2(k)\) does not approach zero at large scales but instead asymptotes to approximately 0.07. This residual plateau arises from small-scale discrepancies that propagate into large scales through the non-linear transformation from density to flux. In the Lyman-\(\alpha\) forest, accurately resolving the pressure-smoothing and thermal-broadening scales is critical for capturing small-scale structure. While HR-HydroSim resolves these scales, residual inaccuracies in the emulator at small scales—due to limited resolution or imperfect modeling—can affect large-scale coherence. Thus, improving small-scale accuracy in the emulator remains essential for reducing decoherence in the large-scale limit.

\subsubsection{Flux Power Spectrum}

The one-dimensional flux power spectrum was computed to assess how well the HydroEmulator captures the clustering properties of the Lyman-$\alpha$ forest. Predictions were benchmarked against high-resolution simulations and observational measurements. Observational data reported by \citet{Day2019}, covering the redshift interval $z = 2.6$–$3.0$ with a mean redshift of $z = 2.83$, and by \citet{walther2019new} at $z = 3.0$, as well as by \citet{Irsic2017} (XQ-100) at $z = 3.0$, were employed. Across the range of scales reliably constrained by the data (up to $k \sim 0.1\ \mathrm{s/km}$), the HydroEmulator’s power spectrum remained consistent with HR-HydroSim results, exhibiting relative differences below 10\%. For scales larger than \( \sim 33\ \mathrm{km/s} \) (i.e., \( k < 3 \times 10^{-2}\ \mathrm{s/km} \), corresponding to \( 2.26\ h/\mathrm{Mpc} \) at \( z = 3 \)), the mean and maximum relative deviations were found to be 1.07\% and 6.67\%, respectively.
HydroEmu maintains high accuracy in the low-$k$ regime—particularly in the mid-$k$ range ($k \sim 10^{-2}\ \mathrm{s/km}$)—where the Lyman-$\alpha$ signal is most informative for cosmological analysis.

\begin{figure}[ht!]
    \centering
    \includegraphics[width=0.95\columnwidth]{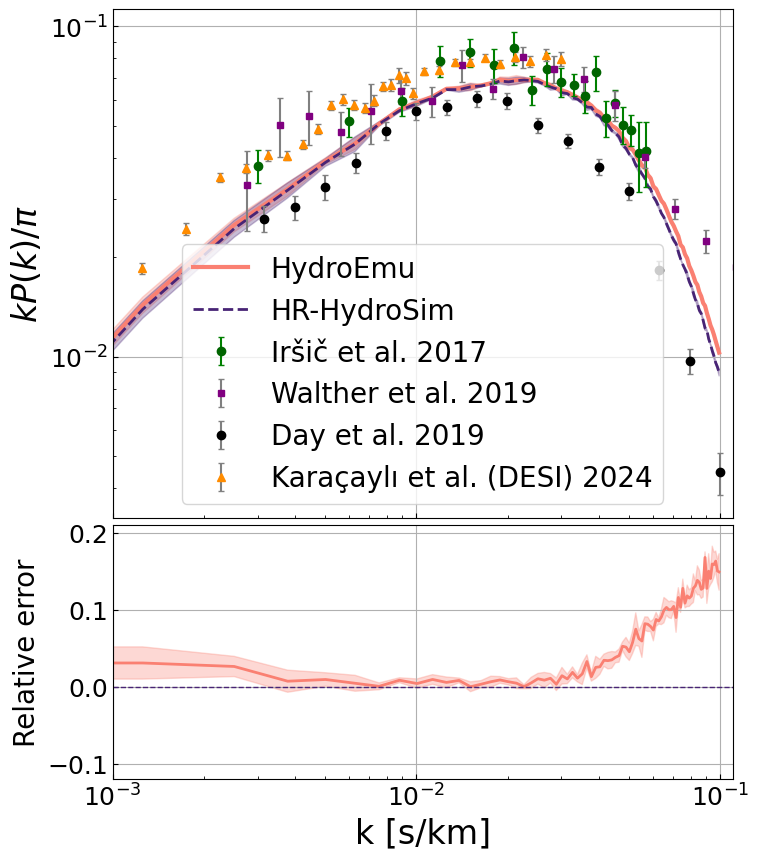}
    \caption{Comparison of the one-dimensional flux power spectra from HydroEmu (solid red) and HR-HydroSim (dashed blue) with observational measurements. Green circles indicate data from \citet{Irsic2017} (XQ-100) at $z = 3.0$; purple squares represent measurements from \citet{walther2019new} at $z = 3.0$; black circles show data from \citet{Day2019} averaged over the redshift range $z = 2.6$–$3.0$; and orange triangles correspond to large-scale measurements from DESI \citep{DESI2024} at $z = 3.0$. The upper panel displays the dimensionless flux power spectrum, $kP(k)/\pi$, while the lower panel shows the relative difference between HydroEmu and HR-HydroSim. Shaded regions denote the standard deviation across simulation realizations.
}
    \label{fig:lya_power_obs}
\end{figure}

\subsubsection{Flux Probability Distribution Function}

\begin{figure}[ht!]
\centering
\includegraphics[width=\columnwidth]{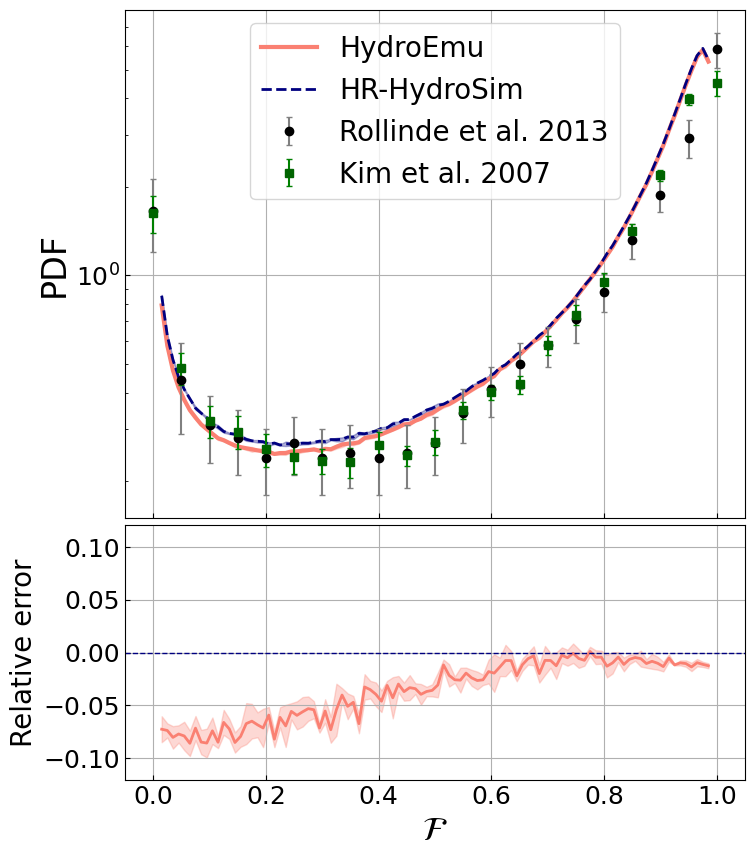}
\caption{Flux PDF from the HydroEmu(solid red) and HR-HydroSim (dashed navy), compared to observational data from \citet{Rollinde2013} and \citep{kim2007improved}. The bottom panel displays the relative error between HydroEmu and HR-HydroSim results, with shaded bands indicating the standard deviation across sightlines.}
\label{fig:flux_pdf}
\end{figure}

The HydroEmulator’s ability to reproduce the statistical distribution of transmitted flux values was further examined using the flux PDF. As illustrated in Figure~\ref{fig:flux_pdf}, HydroEmulator predictions were compared with HR-HydroSim simulations and observational constraints from \citet{Rollinde2013} and \cite{kim2007improved}. The PDF was evaluated over the full range of $\mathcal{F}$ values. Excellent agreement was observed between the HydroEmulator and HR-HydroSim baseline, with relative deviations remaining below 5\% across most of the flux range. The high-flux tail, sensitive to thermal broadening and numerical resolution, showed slightly larger deviations but remained within observational uncertainty. This consistency confirms that the HydroEmulator retains fidelity in capturing small-scale flux statistics relevant to IGM structure.

\subsection{Runtime Comparison}

To quantify the computational efficiency achieved by the deep learning framework, the runtime required for producing high-fidelity outputs was compared against that of a full hydrodynamical simulation using MP-Gadget. The HR-HydroSim was performed on a CPU node, utilizing 4 MPI tasks, each with 14 CPU cores (totaling 56 cores) and 8 GB of memory per core, and it required approximately 267{,}000 seconds to complete. While the LR-HydroSim simulation—run on the same hardware configuration—required approximately 287 seconds, representing a significantly reduced runtime due to its lower resolution. In contrast, the HydroSR and HydroEmu were executed on a GPU node equipped with two A100 GPUs. The HydroSR required only 46 seconds, and the HydroEmu model required 261 seconds to complete their respective tasks. Therefore, the total time to produce outputs using the LR-HydroSim and  deep learning pipeline was reduced to 594 seconds. This corresponds to a speed-up factor of approximately $450\times$ over the full simulation. The significant gain in efficiency illustrates the practicality of employing deep learning methods for accelerating cosmological simulations.

\begin{table}[ht]
\centering
\caption{Runtime and speed-up comparison between simulation and deep learning models.}
\begin{tabularx}{\linewidth}{Xcc}
\hline
\textbf{Method} & \textbf{Hardware} & \textbf{Runtime (s)} \\
\hline
HR-HydroSim & CPU  & 267{,}000 \\
\hline
LR-HydroSim & CPU  & 287 \\
HydroSR & GPU  & 46 \\
HydroEmu & GPU  & 261 \\
Total & GPU \& CPU  & 594 \\
\hline
\end{tabularx}
\label{tab:runtime_comparison}
\end{table}

\section{Discussion}

In this work, HydroEmu was introduced as a deep learning-based field-level emulator capable of reproducing the key baryonic fields relevant to Lyman-$\alpha$ forest modeling—specifically displacement, velocity, and internal energy—at high resolution. Trained on the outputs of cosmological hydrodynamical simulations, HydroEmu reproduces key baryonic fields and flux statistics with 0.1--10\% accuracy across a range of validation metrics, including field morphology, sightline profiles, the temperature--density relation, and flux power spectrum and PDF.
By emulating full 3D fields rather than only summary statistics, our approach enables the generation of synthetic observables (e.g., spectra and tomographic maps), supports a wider range of downstream analyses, and allows for more interpretable connections to physical processes—making it a versatile tool for forward modeling in cosmological inference.

To achieve the final high-fidelity outputs, HydroEmu was used in conjunction with a HydroSR, which enhances the spatial resolution of the low-resolution input fields prior to emulation. This two-stage approach enables the reconstruction of full 3D field-level outputs while remaining highly computationally efficient. When combined, the HydroSR and HydroEmu models reduce the total runtime for generating high-resolution predictions to just over five minutes, corresponding to a speed-up factor of approximately $450
\times$ compared to the original hydrodynamical simulation. Although the resulting model achieves high accuracy and can predict full-box 3D baryonic fields in a fraction of the simulation time, this work represents only the first step in demonstrating the potential of AI to enable fast, scalable alternatives to traditional cosmological baryonic simulations.

This work is just the beginning, and looking ahead, several key directions can extend the impact and applicability of our approach:

\begin{itemize}
    \item \textbf{Direct spectra generation.} One immediate extension is to restructure the inference pipeline such that the emulator generates Ly$\alpha$ absorption spectra on the fly using small chunks of the simulation volume, without requiring the full storage of all particle data. This chunk-wise approach could significantly reduce memory overhead, enabling  much larger simulation volumes. For instance, our current model is trained on 50~cMpc boxes, but the proposed approach could scale to Gigaparsec volumes relevant to large surveys such as DESI.

    \item \textbf{Generalization to multiple cosmologies.} The current model is trained for a fixed cosmological model. A natural next step is to extend the training dataset to include a suite of simulations with varying cosmological parameters. Doing so will allow the model to interpolate baryonic field predictions across a realistic cosmological parameter space, similar in spirit to P1D emulators but at a field level. This would enable more realistic emulation of Ly$\alpha$ flux under varying cosmologies and provide a tool for cosmological parameter inference.

    \item \textbf{Redshift evolution.} Our current emulator operates at a single redshift snapshot ($z=3$). Future versions can incorporate temporal information by either conditioning the model on redshift or jointly training across multiple snapshots. \citep{zhang2025ai} This redshift-aware design would support emulation of the IGM's thermal and dynamical evolution over cosmic time, enhancing the emulator's utility for forward modeling of survey observations and reionization history studies. 
\end{itemize}

Together, these directions will enable HydroEmu to evolve into a general-purpose emulator for cosmological baryonic fields, capable of supporting precision cosmology and large-scale structure analyses at survey scale. Beyond the Lyman-$\alpha$ forest, extensions of this approach could also target other observables sensitive to baryonic physics, such as the Sunyaev–Zel'dovich (SZ) effect \citep{sunyaev1972observations, ade2016planck} or diffuse X-ray background emission from the intracluster medium \citep{rosati2002xray,cappelluti2017chandra}. These multi-wavelength observables provide complementary constraints on the thermal and dynamical state of cosmic baryons and highlight the potential of AI-driven emulators in advancing next-generation cosmological inference.

\begin{acknowledgments}
This research is part of the Frontera computing project at the Texas Advanced Computing Center. Frontera is made possible by NSF award OAC-1818253. TDM acknowledges funding from NSF ACI-1614853, NSF AST-1616168, NASA ATP 19-ATP19-0084, and 80NSSC20K0519. TDM and RACC also acknowledge funding from NASA ATP 80NSSC18K101, and NASA ATP NNX17AK56G. SB was supported by NASA ATP 80NSSC22K1897. This work was also supported by the NSF AI Institute: Physics of the Future, NSF PHY-2020295. 
We also acknowledge the code packages used in this work: The simulations for training and testing were run with \texttt{MP-Gadget} (\url{https://github.com/MP-Gadget/MP-Gadget}). Visualization in this work was performed with the open-source code \texttt{gaepsi2} (\url{https://github.com/rainwoodman/gaepsi2}) and \texttt{plotly} (Inc. 2015). Data and catalogue analysis in this work was performed with open-source software \texttt{PyTorch} \citep{paszke2019pytorch}, \texttt{nbodykit} \citep{hand2018nbodykit}.

\end{acknowledgments}

\section*{Data Availability}

The data and code used in this study are available upon reasonable request to the corresponding author. Simulation outputs generated with \texttt{MP-Gadget} and processed using \texttt{gaepsi2}, \texttt{PyTorch}, and \texttt{nbodykit} can be shared via a data repository upon publication.

\bibliography{ref.bib}
\bibliographystyle{aasjournal}

\end{document}